\documentclass[preprint,times,sort&compress]{elsarticle}

\pdfoutput=1

\journal{ArXiv}

\usepackage{fixltx2e}

\usepackage{url}
\usepackage[breaklinks]{hyperref}

\usepackage{graphicx}

\usepackage{algorithmic}
\usepackage{algorithm}

\usepackage{amsmath}
\usepackage{mathtools}

\usepackage{multirow}
\usepackage{booktabs}
\usepackage{makecell}

\usepackage[format=hang, labelformat=parens]{subcaption}
\usepackage[binary-units]{siunitx}

\usepackage[T1]{fontenc}

\usepackage[acronym,smallcaps]{glossaries}
\newacronym{tfm}{TFM}{Transfer function of the impedance}
\newacronym{icc}{ICC}{Continuity of instrumentation}
\newacronym{orc}{ORC}{Ohmic resistance}
\newacronym{ls1}{LS1}{Long Shutdown 1}
\newacronym{hvq}{HVQ}{High Voltage Qualification}
\newacronym{doc}{DOC}{Dipole Orbit Corrector}
\newacronym{floss}{FLOSS}{Free/Libre and Open Source Software}	
\newacronym{www}{WWW}{World Wide Web}
\newacronym{ict}{ICT}{Information Communication Technologies}
\newacronym{loket}{LOKeT}{Building food market communities with the open source}	
\newacronym{te-mpe-ee}{TE-MPE-EE}{Technology Department, Machine Protection and Electrical Integrity group, Electrical Engineering}
\newacronym{fsf}{FSF}{Free Software Foundation}
\newacronym{mit}{MIT}{Massachusetts Institute of Technology}
\newacronym{cern}{CERN}{European Organization for Nuclear Research}
\newacronym{gpl}{GPL}{General Public Licence}
\newacronym{gplv2}{GPLv2}{GNU General Public License Version 2}
\newacronym{gplv3}{GPLv3}{GNU General Public License Version 3}
\newacronym{lgpl}{LGPL}{GNU Lesser General Public Licence}
\newacronym{osi}{OSI}{Open Source Initiative}
\newacronym{cpl}{CPL}{Common Public License}
\newacronym{epl}{EPL}{Eclipse Public License}
\newacronym{bsd}{BSD}{Berkeley Software Distribution}
\newacronym{agpl}{AGPL}{GNU Affero General Public License}
\newacronym{mpl}{MPL}{Mozilla Public Licence}
\newacronym{bsdnew}{BSD New}{BSD 3-Clause License}
\newacronym{freebsd}{FreeBSD}{BSD 2-Clause License}
\newacronym{mitlicence}{MIT}{MIT licence}
\newacronym{apachelicence}{Apache licence}{Apache License 2.0}
\newacronym{imap}{IMAP}{Internet Message Access Protocol}
\newacronym{pop}{POP}{Post Office Protocol} 
\newacronym{html}{HTML}{HyperText Markup Language}
\newacronym{rss}{RSS}{Rich Site Summary}
\newacronym{odt}{ODT}{Open Document Format}
\newacronym{lamp}{LAMP}{Linux, Apache, MySQL and PHP or Perl or Python}
\newacronym{vcs}{VCS}{Version Control System}
\newacronym{cvcss}{CVCSs}{Centralized Version Control Systems}
\newacronym{dvcs}{DVCs}{Distributed Version Control Systems}
\newacronym{lhc}{LHC}{Large Hadron Collider}
\newacronym{elqa}{ELQA}{Electrical Quality Assurance}
\newacronym{mscas}{MSCAs}{Mobile Social Computing Applications}
\newacronym{rca}{RCA}{Radio Corporation of America}
\newacronym{lep}{LEP}{The Large Electron-Positron Collider}
\newacronym{atlas}{ATLAS}{The Toroidal LHC Apparatus} 
\newacronym{cms}{CMS}{The Compact Muon Solenoid} 
\newacronym{alice}{ALICE}{A Large Ion Collider Experiment}
\newacronym{ssc}{SSC}{Superconducting Super Collider}
\newacronym{totem}{TOTEM}{The Total Cross Section, Elastic Scattering Diffraction Dissociation}
\newacronym{lugos}{LUGOS}{The Linux User Group of Slovenia}
\newacronym{xp}{XP}{Extreme Programming}
\newacronym{ls}{LS}{Long Shutdown}

\makeglossaries

\usepackage[svgnames]{xcolor}
\usepackage{hyperref}

\hypersetup{
    colorlinks = false,
    linkbordercolor = {white},
}

\begin{document}

\begin{frontmatter}

\title{A Conceptual Framework for Supporting a Rapid Design of Web Applications for Data Analysis of Electrical Quality Assurance Data for the LHC}

%\subtitle{Electrical Quality Assurance Data of the LHC}

%\titlerunning{Rapid Prototyping of Web Applications for Data Analysis}

\author[cernaddress]{Matej Mertik}
\ead{matej.mertik@cern.ch}

%\author[cernaddress]{Knud Dahlerup-Petersen}
%\ead{knud.dahlerup-petersen@cern.ch@cern.ch}

\author[aghaddress_eit]{Maciej Wielgosz}
\ead{wielgosz@agh.edu.pl}

\address[cernaddress]{The European Organization for Nuclear Research - CERN, CH-1211 Geneva 23 Switzerland}
\address[aghaddress_eit]{Faculty of Computer Science, Electronics and Telecommunications, AGH University of Science and Technology, Krak\'ow, Poland}

\begin{abstract}
The \gls{lhc} is one of the most complex machines ever build. It is composed of many components which constitute a large system. The tunnel and the accelerator is just one of a very critical fraction of the whole LHC infrastructure. Hardware comissioning as one of the critical processes before running the LHC is implemented during the \gls{ls} states of the macine, where \gls{elqa} is one of its key components. Here a huge data is collected when implementing various ELQA electrical tests. In this paper we present a conceptual framework for supporting a rapid design of web applications for ELQA data analysis. We show a framework's main components, their possible integration with other systems and machine learning algorithms and a simple use case of prototyping an application for Electrical Quality Assurance of the LHC.
\end{abstract}


\end{frontmatter}

%\linenumbers

\section{Introduction}
\label{section:intro}

Visualizing and exploring the data in the Knowledge Discovery in Databases (KDD) process represent important aspects of the process \cite{Fayyad}. Prevalent methods for these aspects are filtering, integrating, visualizing and evaluating the patterns extracted from the data. Since there are many professional and data science toolkits developed today, selecting appropriate ones to use in the domain is not a trivial task. Data science as an emerging field addresses these challenges and offers new directions for organisations to approach and interact with their data. 

The state of software tools for data analysis currently being used by the \glsdesc{elqa} (\gls{elqa}) team at European Organization for Nuclear Research (CERN) is a mixture of high-powered, scientific packages such as LabView alongside simple, web-based, purpose-built tools. These are used to visualize the data and manage the work-flow of ELQA tests that are carried out during the Long Shutdown periods (for example \gls{ls} 1) of the Large Hadron Collider (LHC). This mix of different solutions works well. The LabView tools are used for data acquisition and analyses, which include curve-fitting and Gaussian distribution fitting \cite{Bednarek01, Bednarek02}. The data is acquired using custom-built electronics, designed specifically for making measurements in the LHC tunnel. The web tools are then used to mark magnets or circuits as ''assured'' and fit for the next phase of the LHC commissioning. 

However with the new and new data coming from the hardware commissioning phases of the LHC there is a need for the development of an advanced analysis tool to analyse large amounts of data produced by the hardware commissioning. Under normal circumstances, the data is only used to identify problems with electronics of the LHC after which it remains unanalysed in a database. However, it may be an important mission of the team to investigate the huge number of factors at play that govern the properties of the magnets. Based on this investigation and by gaining further insight more accurate testing on the identified spots can be carried out.

In this paper we present a conceptual framework developed for rapid prototyping of web-based application for exploration and analysis of the \gls{elqa} data. We present practice of merging different software solutions from different software and data scientific communities in order to develop the tool.

The contributions of this paper are: (a) presentation of a framework and its components developed for supporting rapid interactive data analysis; (b) a use case of the prototyping the web application for cleansing and defining patterns in \gls{elqa} data of the LHC. The reminder of the paper is organized as follows. Section 2 discusses the environment and the requirements of \gls{elqa} engineers of the Technology Department, Machine Protection and Electrical Integrity group Electrical Engineering (TE-MPE-EE) section for the data analysis. Section 3 contains the conceptual framework and presents components and properties of the libraries that support it. Experience of merging different software solutions from different software and scientific communities in order to develop the framework are discussed here. Note that some modules of a framework are still under development, however several results have already been achieved. The architecture of application is presented in concluding part of this section. In Section 4 the process of prototyping, a development of a web application for exploring the data in framework is shown on the use case example of the \gls{elqa} for observing capacities of electrical circuits on the LHC. Finally, we give conclusions and areas of further work in Section 5.

\section{Electrical Quality Assurance and Framework Requirements}
\label{sec:1}
The measurement systems in \gls{elqa} group are specialized custom-built systems for measurement of electrical signals of circuit in the LHC tunnel. These are mobile systems that can be rolled on their wheels along the tunnel where a laptop running LabView software then interface with one of these systems at LHC in order to gather data being produced by placing a certain voltage over the selected circuit \cite{Bednarek02}. The specific voltage depends on the type of test and the current machine temperature (cold, warm, cooling down or warming up). 

In the relation database Engineering and Equipment Data (EDMS) all tables in ELQA datasets are connected in a way that many complex relationships can be analysed. For example, you could link a test signal to the position of the magnet in the LHC, the operator(s) who carried out the test, the sector of the magnet, the manufacturer, the number of tests that have been done on it in total, the number of comments that have been made on this test and in some cases the temperature and humidity of the tunnel at the time of testing. This wealth of data that has been stored should provide valuable insight into the tests were carried out.

\subsection{Framework Requirements}
The requirements defined by the \gls{elqa} group for the framework of data analysis were gathered during observation and meetings by designing the use cases. RapidMiner tool was used during this process to generate some analytical models, prototypes of visualisation and overviews of the data on which functionalities were extracted for the framework with the engineers \cite{MertikMJOTE}. Some main characteristic that were identified where: (a) searching for signal's similarities, (b) visualizing trends in the time domain trough the measurements data, (c) developing corresponding data models for different flavours of \gls{elqa} measurements such TP4 (Test Procedure 4), DOC (Dipole Orbit Correctors), MIC (Magnet Instrumentation Check), (d) developing different scenarios of visualisation for measurement of each group of tests.

However, further requirements were added based on the usability and infrastructure at CERN where (a) web environment and (b) the demand for interaction was highlighted as a key point for exploring the data. 

Based on this and the growing usage of python among data miners and data scientists \cite{KDnuggetsNews}, python was chosen as the suitable programming language for the framework. Python also offers a various frameworks for designing a web solutions such as for example Django \cite{Django}, Web2py \cite{Web2py} which were all reconsidered for its use. 

\section{The Conceptual Framework Overview}
\label{sec:2}

\begin{center}
\begin{figure}[h]
  \includegraphics[width=0.75\textwidth]{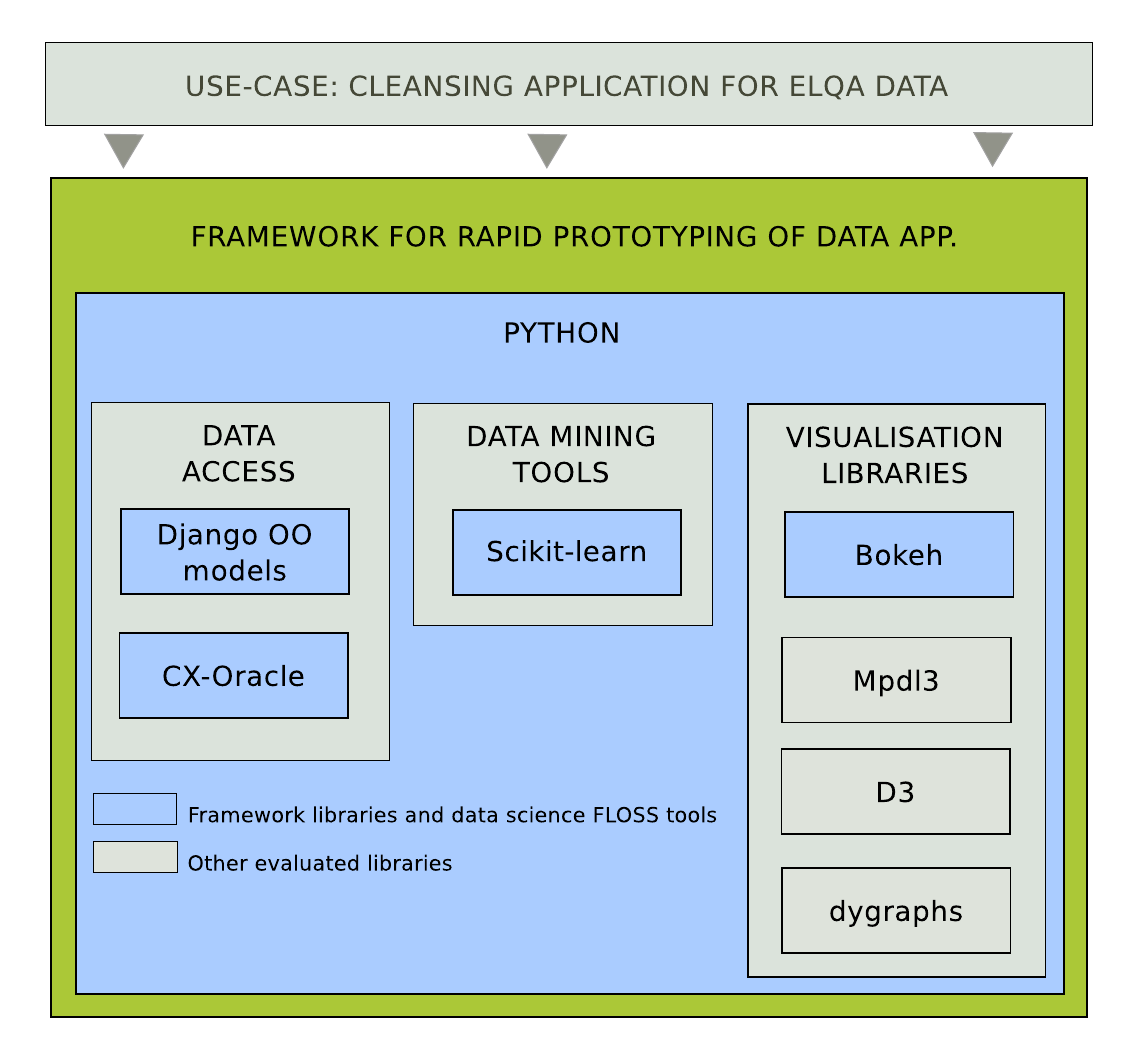}
\caption{The components of the framework for rapid design of data applications}
\label{fig_1}       % Give a unique label
\end{figure}
\end{center}

The framework for rapid design of the data analysis application should in general provide relative simple components for building the applications. There where two main libraries used based on which the concept was defined:  Bokeh Python interactive visualization library that targets modern web browsers for presentation \cite{Bokeh} and was selected based on testing of the visualisation libraries and requirements \cite{lukematej} and Django high-level Python Web framework for Object Oriented paradigm to accessing relational dataset of \gls{elqa} in EDMS database \cite{Django}. Within these two main libraries other depended libraries and components were defined. The relationships between them are shown on the figure \ref{fig_1}.

In the following subsections we are explaining some of the characteristic and experiences of used libraries. We explain the characteristic based on the framework components defined on the phases of the KDD process such as: (a) accessing the data, (b) feature-extraction/preprocessing phase, (c) data mining phase and (d) visualisation and evaluation phase. 

\subsection{Data models}
Accessing the data is addressed with the Object-Relational Mapping that is a frequent topic when it comes to creating Object Oriented programs that require access to a relational database. Django framework handles the mapping with an ease, providing all the functionality of a Structure Query Language (SQL). The architecture of Django is layered.

The bottom layer is the database itself (ELQA dataset in our case). The second bottom layer is an access library. This is responsible for mitigating communication between Python and the database through SQL statements. We used a Python library called cx\_oracle \cite{cxOracle}. This library interface between an Oracle database and any Python program. Django uses this to connect its Model framework to any Oracle database through a small amount of configuration that contains information for authentication among other options.

Above this lies the specific Django database back-end. It can be any kind of database supported by Django (PostgreSQL, MySql, SQLite and Oracle) and there is also the option of writing a customized back-end or using a 3rd-party database back-end \cite{Django}. There are some features that remain unimplemented, one of which had implications for the framework: namely Django only supports finding distinct field values of query sets with the PostgreSQL database back-end. However other libraries can fill in the gaps (NumPy\cite{Numpy} provides a way to find unique values in a list). 

Following the data model comprise the layer above the database back-end. Each one represents a table in the chosen database back-end. The variables of the classes that inherit from the Django base model class map 1-to-1 with the columns from their related database table. The name of the variable is used as the name of the column by default. Each model provides an interface for inserting, selecting, updating and removing rows of the database through a Manager. In this way the framework can elegantly access the Oracle dataset through the Django OO model. These approach is far more agile and adaptable than using SQL statements (which Django are also supports \cite{Django}). However SQL was avoided to ensure that maintainability of the code was maximized.

Above all these there are Django Views models that typically mitigate the connection between HttpRequests and Models. There are not used in our framework as the visualisation is completely embedded within Bokeh library and is explained later.

\subsection{Preprocessors}
Preprocessors are the next concept addressed by the framework and are defined for analytical preprocessing of the signals within defined operators by the ELQA engineers. For the the extraction of signals currently following basic statistics are used: average, minimal value, maximal value, skewness, kurtosis and properties of the linear regression applied to each signal such as slope and standard error. Additional new parameters might be calculated such as for example capacity of the signal that can be used for insight by the end user application.

\begin{center}
\begin{figure}
  \includegraphics[width=1.0\textwidth]{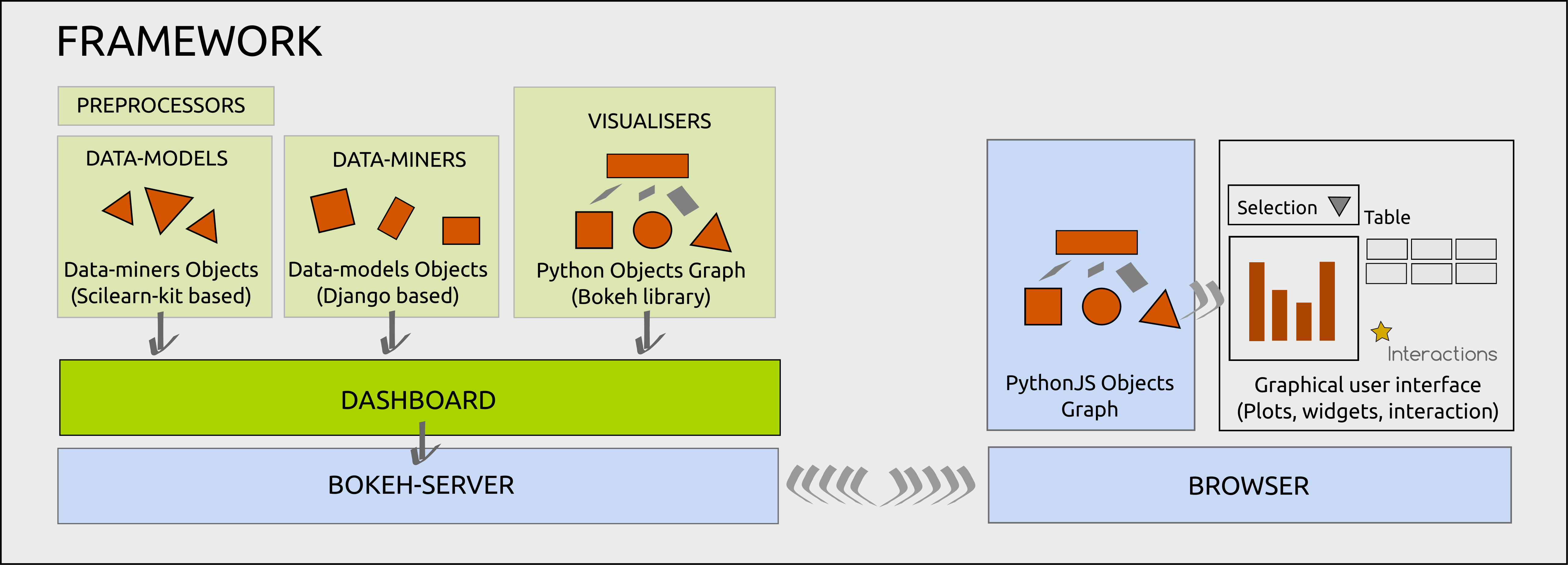}
\caption{The architecture of the framework: the whole application is designed in a Dashboard class where objects from framework are defined, initiated and interconnected within the Bokeh programming paradigm. User interaction between the widgets are defined in the Dashboard and proceeded trough the bokeh-server back-bone architecture where python objects reflects in their java-script relatives}
\label{fig_2}       % Give a unique label
\end{figure}
\end{center}

Scipy and numpy libraries \cite{scipy, Numpy} are used when implementing preprocessor for the framework. Each extended OO Django data model therefore defines an interface where preprocessing techniques could be defined based on the requirements (see preprocessors objects on architecture of the framework \ref{fig_2} respectively).

\subsection{Data-miners}
There are many solutions and tools for data analysis available as open-source or commercial packages. Some of the most used tools from the domains of Data Mining, Analytics, Big Data, and Data Science can be found on KDnuggets network maintained by a group of professionals and researchers \cite{KDnuggets}. Scikit-learn, a Machine Learning library written in Python was appropriate and was used for the implementation of the analysis \cite{scikit-learn}. It is built on NumPy, SciPy, and matplotlib and it licensed under BSD open source license. 

The data-miner layer defines general class analyser in the framework. Currently two different clustering algorithms are integrated from the skilearn-kit and are used in the framework for searching similarities between the signals K-means clustering \cite{Kmeans}, a popular type of cluster analysis and DBScan clustering, which is a density-based spatial clustering of data with noise proposed by Martin Ester, Hans-Peter Kriegel, Jorg Sander and Xiaowei Xu in 1996 \cite{Ester96adensity-based}

The analyser might be natural extended with other techniques from scikit-learn or other libraries. An experiment work of deep learning building model, a model based on Deep Learning algorithms of LSTM and GRU for facilitating anomaly detection in LHC superconducting magnets \cite{Wielgosz2016UsingRecurrent} is under evaluation. Here high resolution data available in the MTF (Magnet Test Folder) dataset of Post-Mortem (PS) system of the LHC is used to train a set of models and chose the best possible set on their hyper-parameters. Further techniques of deep learning are reconsidered for developement in applications such as efficient feature extraction techniques. 

\subsubsection{Boosting the feature-extraction-based approach with Deep Learning methods}
There is a common opinion that a Deep Learning approach is in an opposition to methods base on feature extraction paradigm \cite{Haibo, Peng, Islam, Wielgosz2016Using} in terms of their application capabilities. Consequently, it is often claimed that it is recommended to use Deep Learning methods instead of laborious manual feature preparation and extraction whenever it is possible. This however is not alway a case. Furthermore, there are applications where those two approach may coexist since then supplement each other in many respects. 

Since the approach based on feature extraction is tedious and time consuming it is important when and where to apply it in order to get the best performance of the designed system. This in turned may be determined by using Deep Leaning algorithms to determine sensibility of employing more demanding feature-based methods. 

For instance, given a substantial volume of data which is expected to reflect anomalous behavior of magnets \cite{Wielgosz2016UsingRecurrent} and having no prior knowledge on a its profile whatsoever, one may use Recurrent Neural Networks (RNN) to preliminarily examine the data. Such an investigation may provide information regarding an extent to which the voltage time series deviant from their standard values by studying Root Mean Sqare Error (RMSE) of the results. This in turn can justify an effort to be invested in employing more labour-intensive feature-extraction based approach. The preliminarily examination may also provide insight regarding a character of the expected anomalies which in turn allows for using appropriate machine learning method such as Random Forest, K-mmeans etc. 

A key here is an acquisition of the prior knowledge about the data with fast to train and use Deep Learning methods. Then and advantage may be taken of this information to adopt adequate feature-based machine learning algorithms to profoundly examine the data in the second step.

\subsection{Visualisers}
Visualisers are on of key components for prototyping data analysis applications and serve for for user interaction, visualisation and the exploration of the data. When searching for appropriate tools in communities many frameworks in different programming languages were reconsidered. Due to the required architecture (see other evaluated libraries on the \ref{fig_1}) Bokeh was selected after carefully studying the appropriate available solutions due to its ability to act as a web server in order to expose a fully-fledged data visualization and analysis tool to multiple users \cite{lukematej}. 

At \gls{elqa} section the visualization of the \gls{elqa} signals was currently possible only by using the existing web tools implemented as PHP scripts that are run by an Apache web server \cite{Apache}. The results of such visualization are static plots rendered as PNG images. These visualizations are visualization of raw measurements where there is no option to filter or analyse them. The import to specific data tools should be done for further analysis in specific tools. The solution available in the group is therefore not expandable for modern web visualisation at present.  

The developed framework presented here extends these drawback of currently used tool with interactivity on the fly and focuses on maximizing usefulness and extendibility based on Bokeh library. In the framework the data model is separated from the analyser model and visualization model based on the input changes of the user. In that manner purpose request to analyse data on the server is supported in a format that allows rendering the data by JavaScript on the client which is part of the Bokeh library.

\subsection{Framework architecture}
\label{sec24}

\subsubsection{Bokeh design}
Bokeh Python library is capable of rendering data into interactive, Canvas-based, web plots \cite{Bokeh}. Bokeh is being actively developed. This can be seen from its release notes and Github page \cite{GithubBokeh}. Receiving thousands of downloads per month, having 83 contributors and having 193 commits to master (at the time of writing), Bokeh is certainly alive and being improved by a growing community. The disadvantages of this are usual disadvantages of FLOSS (Free Libre Open Source Software) software with potential bugs and features to be desired, such as plot legends that can be updated dynamically \cite{BokehUpdate} for example. The architecture of Bokeh is defined in a way such that data can be sent and received to and from the client browser for the duration that the user stays on the page. This allows for session-long interactivity and hence the \'analyze, visualize, mine\' cycle. Callback functions can be defined in Python code that are called when certain JavaScript events occur such as the selection of points on a graph. Bokeh comes with UI elements such as selection boxes that can also be linked to Python functions, which are called when a selection is made. This allows for data to be re-analysed without any UI data being resent to the browser unnecessarily (as HTML). For example, a user can select some points with a selection box and the plot below responds and displays data related to those points, such as the signal that was used to generate that point, if the point represents part of a feature vector of that signal. 

Bokeh in general defines Documents, which are objects that can store the data and the layout data required to render a user interface with the currently selected data. All of this data is stored on a Bokeh server within a chosen back-end. In the case of the presented framework no persistent storage is necessary and the data is stored in memory. Other Bokeh options include a Redis \cite{redis} database.

\subsubsection{Architecture}

The framework for rapid design is designed based on the Bokeh library components where Bokeh server is built on top of Flask \cite{bokehflask}. This concept present the backbone in which components of data models, preprocessors and data-miners are inserted. These blocks meet in the Dashboard subclass where a customized Model and an Analyser are defined on which then bokeh's widget and plots can be manipulated. The Dashboard is the main entrance point of the framework where effective developing of a data analysis application is done. In such a way the amount of work for providing an interactive plots in the browser with widget is improved as whole application is designed within the Dashboard class. This results in the complex interactions between multiple plots on large datasets where corresponding widgets and analysers can be used in reasonable simplified way that enables interactive filtering and some of analysis of the data on the fly. On figure \ref{fig_3} the framework architecture is presented. On top of it the web application for data analysis are then designed. The use case of the ELQA application prototyping is presented in the details.

\section{Use case: building a cleansing application for ELQA data}
\label{sec:3}
In \gls{elqa} section there are needs for reporting the conditions of the circuits in the machine by implemented measurements through the hardware commissioning procedure. The quality of the \glsdesc{hvq} (\gls{hvq}) measurements of the circuits can change from phase to phase in the hardware commission and that can not be seen directly. The capacity of each circuits of different magnets has an important value when observing the health of the each circuit. In the current situation such an inquires could be implemented trough the complex SQL statements and then processed in the exterior tools for further analysis, however there are at least three drawbacks that this is not so simple task: (a) due to the huge amount of data, this is often not convenient, (b) this kind of procedure requires complex SQL inquirements, (c) the data in the dataset are often not cleansed and preprocessed (the important is results of the measurement at the first place when circuit was marked with "assured", however there are also some measurements recorded that were performed for example for only testing purposes as all activities on the machine are recorded into the dataset). Moreover, design of such special application within the existed software in the group would need a large portion of programming in PHP and Javascript of some of the basic things for example plotting the data. Even if this could be done, the life-cycle of such small special design would not be long-term oriented. 

\begin{center}
\begin{figure}[h!]
  \includegraphics[width=0.9\textwidth]{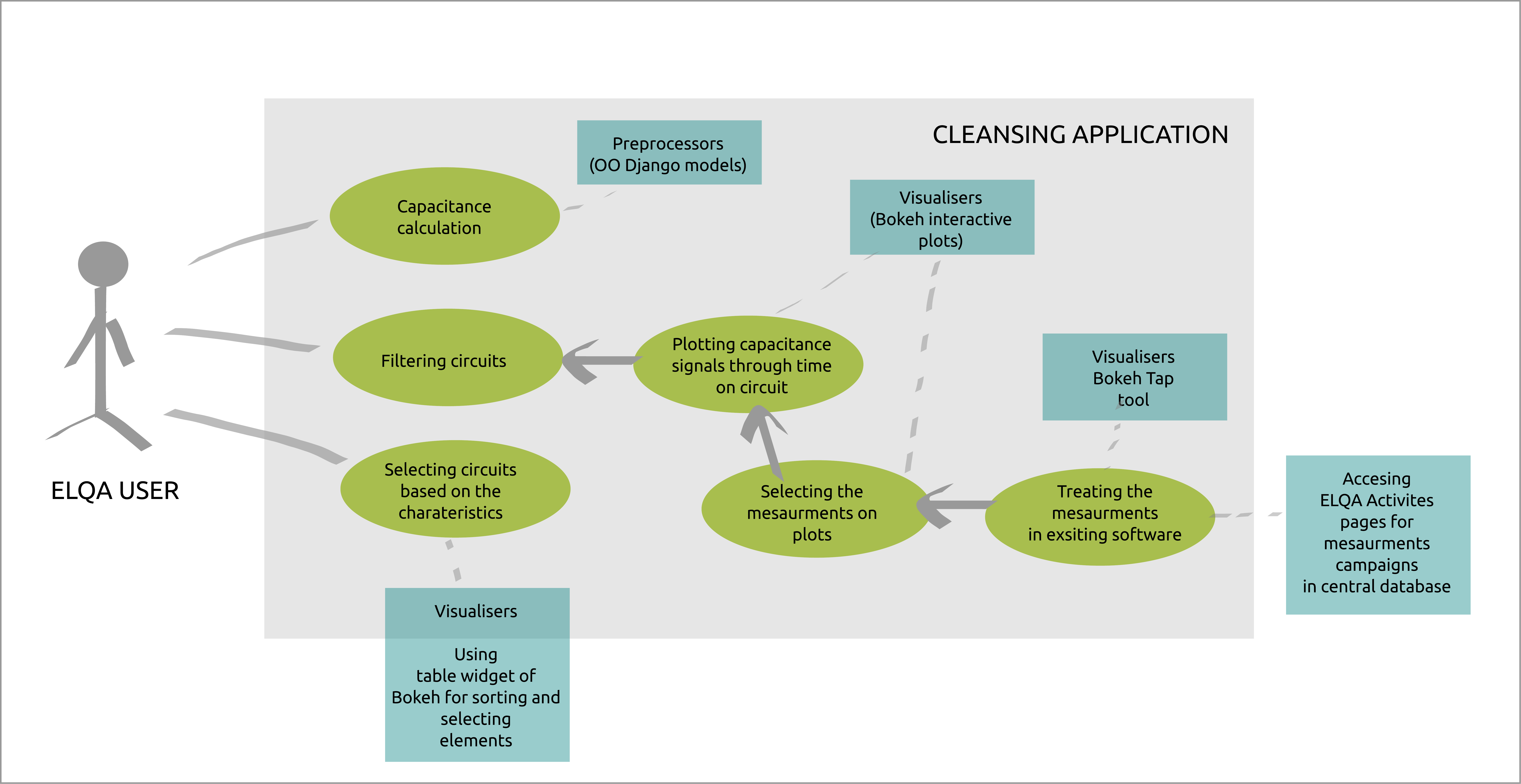}
\caption{The use case for implementing cleansing application. Some of the objects used in Dashboard are spotted in the comments.}
\label{fig_3}       % Give a unique label
\end{figure}
\end{center}

With presented framework it was possible to design this kind of application in a relative short period of time. The requirements for the application were following: (a) providing right calculation of capacitance for each circuits, where missing data should be treated appropriate (preprocessors), (b) filtering efficient selection of the circuits by some of their properties (use of the table widget and some basic statistic), (c) plotting capacitance of the circuits trough time and observing trends in the measurements, (d) being able to spot a measurement with suspicious values and check all the data in the software used for recording the measurements. In the figure \ref{fig_3} a use case diagram for user requirements is presented.

\begin{center}
\begin{figure}[h!]
  \includegraphics[width=0.825\textwidth]{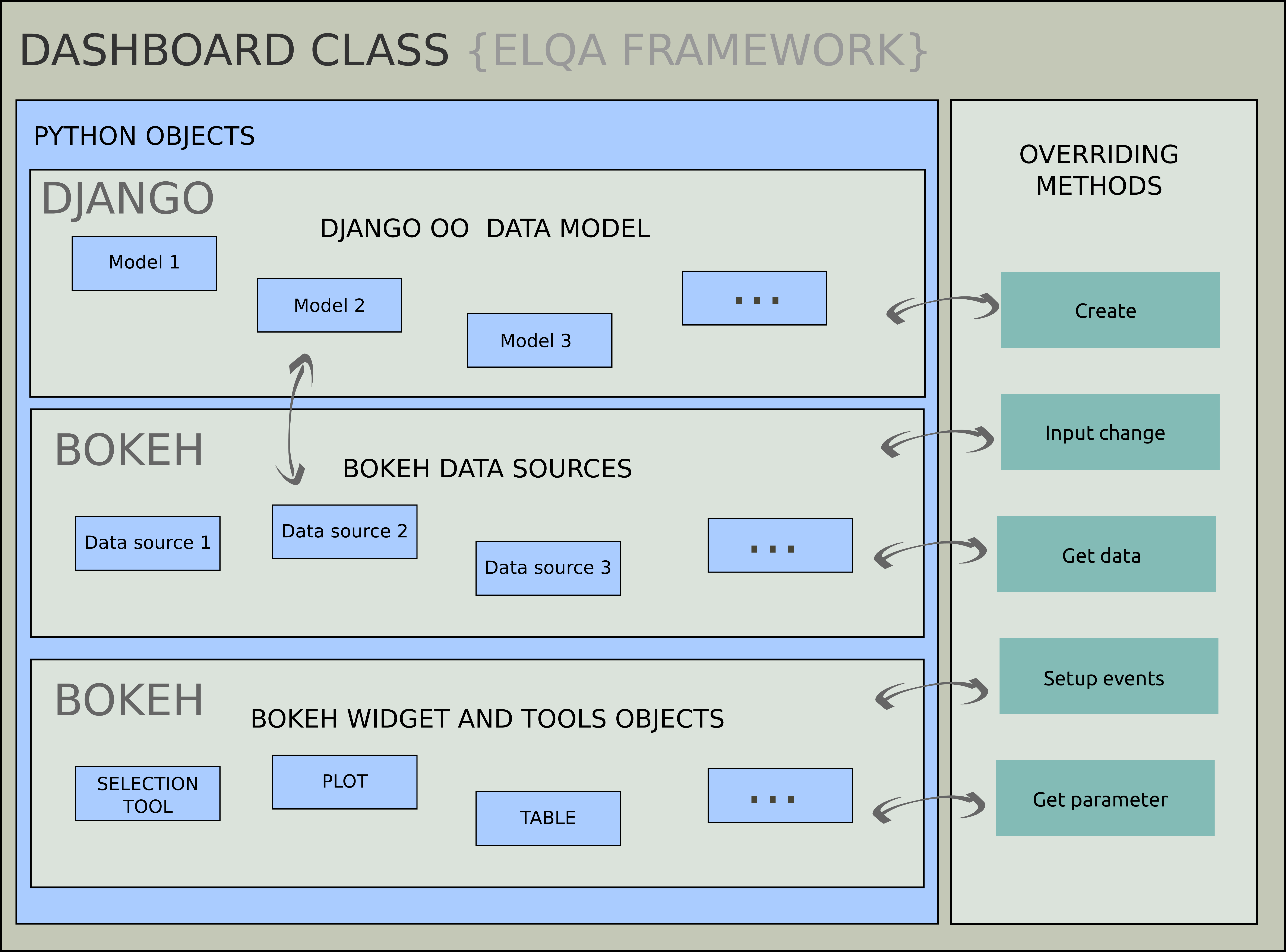}
\caption{Objects and methods in the Dashboard.}
\label{fig_4}       % Give a unique label
\end{figure}
\end{center}

The whole application was designed within the Dashboard class. The Dashboard class consists from  necessary class methods:
(a) create - here object are created such as plots, selection boxes, tables, the data sources and bokeh tools which are initialized with the widgets, (b) setup events - is the method where we define attachments of the user events to the value property of the selected widget, (c) input change - is responsible for updates of the data sources and widgets when the interaction is happening, (d) get data - within this method we defined data sources e.g. which data from the data model we want to plot on the widgets, (e) get parameter - is the method that returns data values for selection box widgets when filtering happens. Figure \ref{fig_4} shows the Dashboard class structure.

\begin{center}
\begin{figure}[h!]
  \includegraphics[width=1.0\textwidth]{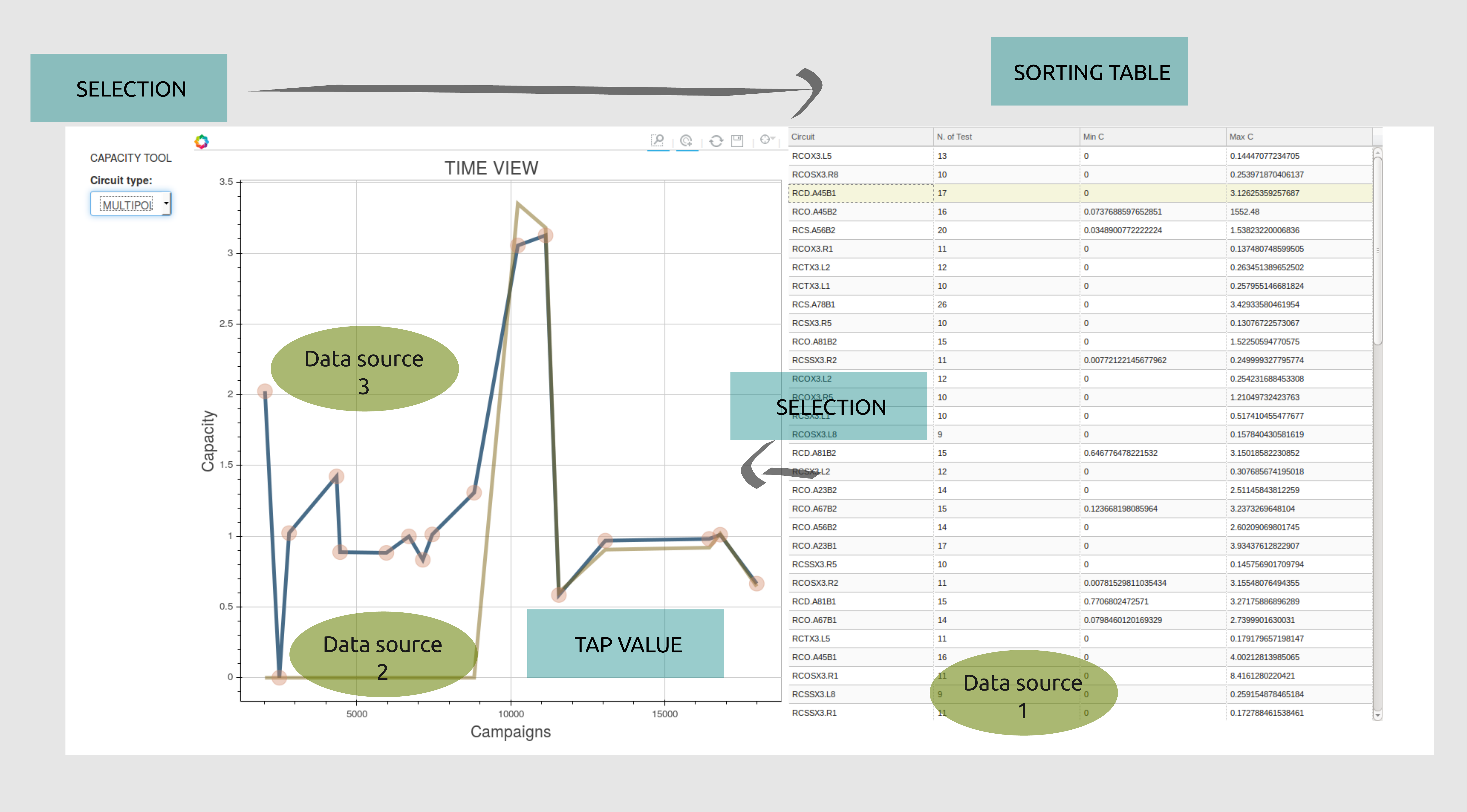}
\caption{Graphical user interface of the application. Shown are interaction for user and data sources associated with the widgets.}
\label{fig_5}       % Give a unique label
\end{figure}
\end{center}

On the figure \ref{fig_5} the cleansing application is shown. Defined are interaction in the application and data sources from the Dashboard diagram. You can see that in this way it is possible to filter data based on the type of circuits. When filtered, they are presented in the table with some characteristics of on each circuit. User can then interact with a table and sort the circuit based on some statistics defined by the \gls{elqa}. When user select the circuit, central plot is refreshing with the all values of capacitance from two different measurements done on each circuit. User can then select particular measurement on the plot and with a tap tool access to the ELQA activities pages for measurements campaigns in central database, where he can then correct the values of the test in order to check and perform the cleansing of the dataset with those measurements that were recorded, if necessary. With this application it was also possible to discover those test which were successfully done, however have a noise within that was produced by the design of TP4 systems and was not treated during the previous campaigns. Corrections on TP4 are now successfully done.

\section{Conclusions and further work}
\label{sec:4}
In this research, we have addressed the architecture of conceptual framework for rapid prototyping the application of the data analysis embedded in the web. We have identified requirements for such a framework, the available open source and scientific tools for data mining, web visualization and data access and we have addressed the non trivial task of selecting an appropriate set of tools for integrating them into the conceptual framework. Our framework allows to facilitate the development of the scientific data analysis web applications with interactive abilities and integrations. We have then presented a prototype of the application for \gls{elqa} data of the LHC's electrical circuits, which uses the framework and have discussed some of the aspects that need to be considered.   

As python was selected as the implementing language, different python libraries were investigated to fit the requirements. Two main python components libraries were selected - Scikit-learn, a Machine Learning library in Python for data analysis and Bokeh, a Python library capable of rendering data into interactive, Canvas-based, web plots. Although importance of these two main components, many additional parameters had to be addressed when integrating libraries together. For accessing the data according to the CERN ELQA database standards and policies, the cx-oracle library was used. On top of this, the Django ORM model was used for achieving access to the relational database using the OO paradigm.

Main components of the conceptual framework are Dashboards which presents a common point for merging the data from the database with the appropriate analysis algorithms from the Scikit-learn and Bokeh libraries. A use case of rapid application prototyping is based on the Dashboard component. However it should be stressed that in order to achieve the integration of libraries in the Dashboard many issues had to be resolved in relation to the Bokeh library. Some contributions were made to the library from the TE-MPE-EE department such as a fix for a bug involving decimal format numbers and the ability to run a HTTPS Bokeh server. For achieving an interaction between the widget some additional callbacks were also defined in JavaScript.

Further work will consist of developing testing environment for the framework and further simplification of the components in the Dashboards. Since the relation between components in Dashboard are well defined, the other direction for further work will be concerned such as the development of metatool for design of graphical user interfaces for generating the Dashboards as the prototyping the application for quench prediction based on the research of deep learning models within the group.

\section*{Acknowledgment}
This research was supported by TE-MPE-EE group at CERN and is being tested to prototype data applications by the ELQA team. The authors would like to thank the entire ELQA team at CERN for their persistent help throughout the ongoing development process and to Knud Dahlerup-Petersen for his supervision and valuable comments.

\nocite{*}

\bibliographystyle{elsarticle-num}
%\bibliography{bibliography,authors-published,authors-unpublished}
\bibliography{elqa_framework}

\begin{thebibliography}{10}
\expandafter\ifx\csname url\endcsname\relax
  \def\url#1{\texttt{#1}}\fi
\expandafter\ifx\csname urlprefix\endcsname\relax\def\urlprefix{URL }\fi
\expandafter\ifx\csname href\endcsname\relax
  \def\href#1#2{#2} \def\path#1{#1}\fi

\bibitem{Fayyad}
U.~Fayyad, G.~Piatetsky-Shapiro, P.~Smyth,
  \href{http://doi.acm.org/10.1145/240455.240464}{The kdd process for
  extracting useful knowledge from volumes of data}, Commun. ACM 39~(11) (1996)
  27--34.
\newblock \href {http://dx.doi.org/10.1145/240455.240464}
  {\path{doi:10.1145/240455.240464}}.
\newline\urlprefix\url{http://doi.acm.org/10.1145/240455.240464}

\bibitem{Bednarek01}
M.~Bednarek, D.~Bozzini, V.~Chareyre, J.~Ludwin, N.~Lasheras, G.~D'Angelo,
  R.~Mompo, Elqa qualification of the superconducting circuits during hardware
  commissionning, Tech. rep., CERN (2013).

\bibitem{Bednarek02}
M.~Bednarek, J.~Ludwin, Software tools for electrical quality assurance in the
  lhc, in: Proceedings of ICALEPCS2011, Grenoble, France, 2008, pp. 993--995.

\bibitem{MertikMJOTE}
M.~Mertik, M.~Bednarek, K.~Dahlerup-Petersen,
  \href{https://doi.org/10.1520/JTE20150202}{Constructing and testing data
  models for lhc electrical quality assurance}, Journal of Testing and
  Evaluation 44~(2) (2016) 1027--1034.
\newblock \href {http://dx.doi.org/10.1520/JTE20150202}
  {\path{doi:10.1520/JTE20150202}}.
\newline\urlprefix\url{https://doi.org/10.1520/JTE20150202}

\bibitem{KDnuggetsNews}
G.~Piatetsky, R leads rapidminer, python catches up, big data tools grow, spark
  ignites,
  \url{http://www.kdnuggets.com/2015/05/poll-r-rapidminer-python-big-data-spark.html}
  (2015).

\bibitem{Django}
D.~S. Foundation, Django documentation,
  \url{https://docs.djangoproject.com/en/1.8/ref/databases/} (2015).

\bibitem{Web2py}
W.~community~of professionals, web2py,
  \url{http://www.web2py.com/init/default/documentation} (2015).

\bibitem{Bokeh}
C.~Analytics, Bokeh reference guide,
  \url{http://bokeh.pydata.org/en/latest/docs/reference.html} (2015).

\bibitem{lukematej}
L.~Barnard, M.~Mertik, Article: Usability of visualization libraries for web
  browsers for use in scientific analysis, International Journal of Computer
  Applications 121~(1) (2015) 1--5, full text available.

\bibitem{cxOracle}
Sourceforge, cx\_oracle, \url{http://sourceforge.net/projects/cx-oracle/}
  (2015).

\bibitem{Numpy}
S.~van~der Walt, S.~Colbert, G.~Varoquaux, The numpy array: A structure for
  efficient numerical computation, Computing in Science Engineering 13~(2)
  (2011) 22--30.
\newblock \href {http://dx.doi.org/10.1109/MCSE.2011.37}
  {\path{doi:10.1109/MCSE.2011.37}}.

\bibitem{scipy}
E.~Jones, T.~Oliphant, P.~Peterson, et~al.,
  \href{http://www.scipy.org/}{{SciPy}: Open source scientific tools for
  {Python}}, [Online; accessed 2017-02-04] (2001--).
\newline\urlprefix\url{http://www.scipy.org/}

\bibitem{KDnuggets}
G.~Piatetsky, \url{http://www.kdnuggets.com} (2015).

\bibitem{scikit-learn}
F.~Pedregosa, G.~Varoquaux, A.~Gramfort, V.~Michel, B.~Thirion, O.~Grisel,
  M.~Blondel, P.~Prettenhofer, R.~Weiss, V.~Dubourg, J.~Vanderplas, A.~Passos,
  D.~Cournapeau, M.~Brucher, M.~Perrot, E.~Duchesnay, Scikit-learn: Machine
  learning in {P}ython, Journal of Machine Learning Research 12 (2011)
  2825--2830.

\bibitem{Kmeans}
J.~A. Hartigan, M.~A. Wong,
  \href{http://www.jstor.org/stable/2346830}{Algorithm as 136: A k-means
  clustering algorithm}, Journal of the Royal Statistical Society. Series C
  (Applied Statistics) 28~(1) (1979) pp. 100--108.
\newline\urlprefix\url{http://www.jstor.org/stable/2346830}

\bibitem{Ester96adensity-based}
M.~Ester, H.~peter Kriegel, J.~S, X.~Xu, A density-based algorithm for
  discovering clusters in large spatial databases with noise, AAAI Press, 1996,
  pp. 226--231.

\bibitem{Wielgosz2016UsingRecurrent}
M.~Wielgosz, A.~Skocze\'n, M.~Mertik,
  \href{http://arxiv.org/abs/1611.06241}{Using {LSTM} recurrent neural networks
  for detecting anomalous behavior of {LHC} superconducting magnets}, CoRR
  abs/1611.06241.
\newline\urlprefix\url{http://arxiv.org/abs/1611.06241}

\bibitem{Haibo}
H.~He, E.~A. Garcia, {Learning from imbalanced data}, IEEE Transactions on
  Knowledge and Data Engineering 21~(9) (2009) 1263--1284.
\newblock \href {http://dx.doi.org/10.1109/TKDE.2008.239}
  {\path{doi:10.1109/TKDE.2008.239}}.

\bibitem{Peng}
P.~Zhang, X.~Zhu, L.~Guo, {Mining Data Streams with Labeled and Unlabeled
  Training Examples}, in: 2009 Ninth IEEE International Conference on Data
  Mining, IEEE, IEEE, Miami, USA, 2009, pp. 627--636.
\newblock \href {http://dx.doi.org/10.1109/ICDM.2009.76}
  {\path{doi:10.1109/ICDM.2009.76}}.

\bibitem{Islam}
M.~K. Islam, F.~Jahan, J.-H. Min, J.-H. Baek, {Object classification based on
  visual and extended features for video surveillance application}, in: Control
  Conference (ASCC), 2011 8th Asian, IEEE, Kaohsiung, Taiwan, 2011, pp.
  1398--1401.

\bibitem{Wielgosz2016Using}
M.~Wielgosz, M.~Pietro\'n, K.~Wiatr, Using spatial pooler of hierarchical
  temporal memory for object classification in noisy video streams, in: 2016
  Federated Conference on Computer Science and Information Systems (FedCSIS),
  2016, pp. 271--274.

\bibitem{Apache}
T.~A.~S. Foundation, Apache http server project,
  \url{http://httpd.apache.org/ABOUT_APACHE.html} (2015).

\bibitem{GithubBokeh}
C.~Analytics, Bokeh release,
  \url{http://bokeh.pydata.org/en/latest/docs/release\_notes.html} (2015).

\bibitem{BokehUpdate}
C.~Analytics, Bokeh issues, \url{https://github.com/bokeh/bokeh/issues/589}
  (2015).

\bibitem{redis}
Redislabs, Redis documentation, \url{http://redis.io/documentation} (Accessed
  August 2015).

\bibitem{bokehflask}
Bokeh, Bokeh documentation,
  \url{http://bokeh.pydata.org/en/0.10.0/docs/user_guide/server.html} (Accessed
  August 2016).

\bibitem{Python1995}
G.~van Rossum, \href{https://www.python.org/}{Python tutorial}, Tech. rep.,
  Centrum voor Wiskunde en Informatica (CWI) (1995).
\newline\urlprefix\url{https://www.python.org/}

\bibitem{LHC}
CERN, Large hadron collider - offcial page,
  \url{http://home.web.cern.ch/topics/large-hadron-collider} (2015).

\bibitem{LHCBulettin}
C.~Bulletin, The lhc goes 3g,
  \url{https://cds.cern.ch/journal/CERNBulletin/2013/08/News\%20Articles/1514604}
  (2015).

\end{thebibliography}

\end{document}